# On the Prony Series Representation of Stretched Exponential Relaxation

John C. Mauro[*] and Yihong Z. Mauro

*Department of Materials Science and Engineering*

*The Pennsylvania State University*

*University Park, Pennsylvania 16802, USA*

[*]*Corresponding Author:* jcm426@psu.edu

**Abstract:** Stretched exponential relaxation is a ubiquitous feature of homogeneous glasses. The stretched exponential decay function can be derived from the diffusion-trap model, which predicts certain critical values of the fractional stretching exponent, $\beta$. In practical implementations of glass relaxation models, it is computationally convenient to represent the stretched exponential function as a Prony series of simple exponentials. Here, we perform a comprehensive mathematical analysis of the Prony series approximation of the stretched exponential relaxation, including optimized coefficients for certain critical values of $\beta$. The fitting quality of the Prony series is analyzed as a function of the number of terms in the series. With a sufficient number of terms, the Prony series can accurately capture the time evolution of the stretched exponential function, including its "fat tail" at long times. However, it is unable to capture the divergence of the first-derivative of the stretched exponential function in the limit of zero time. We also present a frequency-domain analysis of the Prony series representation of the stretched exponential function and discuss its physical implications for the modeling of glass relaxation behavior.

**I. Introduction**

Glasses are inherently nonequilibrium materials that spontaneously relax toward the metastable supercooled liquid state [1]-[14]. While effectively frozen at low temperatures, this relaxation process is accelerated during heat treatment. Glass relaxation is known as one of the most difficult problems in condensed matter physics [15] and one of the most important grand challenges for glass science [16]-[18]. The complicated nonexponential character of the relaxation process has been highlighted by Dyre as one of the ten themes of viscous liquid dynamics [19] and is a critical problem for the statistical mechanics of the glassy state [20].

In addition to its scientific importance, relaxation is of great practical relevance for a variety of high-tech applications of glass. For example, the performance of liquid crystal display glass is governed by volume relaxation, i.e., compaction [21]. This is especially important for high performance display applications, where smaller pixel sizes lead to tighter requirements on glass compaction to avoid misalignment of pixels in the final display [22]. Moreover, the ion exchange temperature used to chemically strengthen glass is fundamentally limited by stress relaxation in the glass [23]. Relaxation also has a large impact on the kinetics of the mutual diffusivity process and the magnitude of the resulting compressive stress from ion exchange [24]. The intrinsic damage resistance of certain glass compositions is also a strong function of its degree of disequilibrium [25]-[27]. The attenuation of optical fiber for telecommunication is dominated by Rayleigh scattering, which is a complicated non-monotonic function of the thermal history of the glass and can be optimized through proper control of the glass relaxation process [28]-[30]. Such non-monotonic decay has also been observed with respect to the relaxation of enthalpy fluctuations in glass [31]. Finally, the coefficient of thermal expansion (CTE) of ultra-low expansion glass depends on its thermal history. During glass relaxation, the zero-crossing point of the CTE curve can shift to different temperatures [32]. Hence, control of the glass transition and relaxation processes is essential for obtaining the desired CTE curve.

The glass relaxation phenomenon itself is an exponentially complex problem not soluble by conventional algebraic means [33]. The scientific study of relaxation dates back to the pioneering work of Kohlrausch [34]-[35] in 1854, who proposed the following empirical equation to describe the relaxation of residual charge on a Leyden jar:

$$g(t) = \exp\left[-(t/\tau)^\beta\right], \tag{1}$$

where $t$ is time. This is commonly known as the "stretched exponential" function, having two free parameters: the relaxation time, $\tau$, and the stretching exponent, $\beta$. The stretching exponent satisfies $0 < \beta \leq 1$, where the upper limit of $\beta = 1$ gives simple exponential decay and smaller values of $\beta$ indicate a nonexponential relaxation process with a "fat tail" at long time. While several types of distributions exhibit fat tails [36]-[38], the stretched exponential function has been shown to provide a universal description of the relaxation behavior of homogeneous glasses, independent of glass composition and the type of probe used to measure the relaxation [33]. The stretched exponential function is also commonly used to describe the relaxation and retardation functions of linear viscoelastic media [39]-[40].

For well over a century since its introduction by Kohlrausch, the physical origin of the stretched exponential function had been considered as one of the oldest unsolved problems in physics [33]. The first physical model to reproduce the stretched exponential functional form was published by Grassberger and Procaccia in 1982 based on the diffusion of excitations to randomly distributed traps that annihilate these excitations [41], where the distribution of relaxation times gives rise to the stretched exponential function. In 1994, this diffusion-trap model was extended by Phillips [42], who showed that the value of the dimensionless stretching exponent can be derived based on the effective dimensionality of the relaxation pathways in the configurational phase space, $d^*$. According to the Phillips model, the stretching exponent is related to $d^*$ by [42]-[45]:

$$\beta = \frac{d^*}{d^* + 2}. \tag{2}$$

The dimensionality of the relaxation pathways can be expressed as $d^* = fd$, where $d$ is the dimensionality of the system and $f$ is the fraction of relaxation pathways activated for the particular process under study. For a three-dimensional structural glass with $d = 3$ and all relaxation pathways activated ($f = 1$), a stretching exponent of $\beta = 3/5$ is immediately obtained. Likewise, relaxation of a two-dimensional film ($d = 2$) with all pathways active ($f = 1$) follows $\beta = 1/2$. A fractal dimensionality is obtained if one assumes an equipartitioning of the relaxation channels into long- and short-range contributions. If we assume that the rate of entropy production is maximized, then there must be a balancing of the long- and short-range contributions. If all of the short-range relaxation processes have finished and only the long-range contributions remain, then $f = 1/2$ and a fractal dimensionality of $d^* = 3/2$ is obtained. With $d^* = 3/2$, a stretching exponent of $\beta = 3/7$ is found for relaxation governed solely by long-range interactions [46].

The success of the diffusion-trap model for stretched exponential relaxation in microscopically homogeneous systems has been confirmed by over 50 examples from the literature [33]. More recently, a definitive set of experiments by Potuzak et al. [47] showed that homogeneous glasses manufactured by the proprietary fusion draw process exhibit a bifurcation of the stretching exponent into $\beta = \{3/5, 3/7\}$ for stress relaxation and structural relaxation processes, respectively. In other words, the same glass composition at the same temperature can exhibit different stretching exponents depending upon the type of relaxation under study. A stress relaxation experiment under an applied load has both short- and long-range relaxation pathways activated and therefore follows stretched exponential decay with $\beta = 3/5$. In the absence of an applied load, the same glass exhibits structural relaxation with $\beta = 3/7$, since only the long-range relaxation pathways are activated. The value of $\beta = 3/7$ for structural relaxation was confirmed in an experiment by Welch et al. [48], who measured the structural relaxation of Corning® Gorilla® Glass over a period of 1.5 years at room temperature, more than 600°C below its glass transition temperature. These critical values of the stretching exponent have also recently been obtained through atomic scale simulations in a series of papers by Yu et al. [49]-[51] The bifurcation of the stretching exponent into $\beta = \{3/5, 3/7\}$ has been demonstrated even outside the traditional realm of physics, e.g., in a study of the statistical distributions of citation chains over time in the scientific literature [52].

In the modeling of glass relaxation behavior, it is mathematically convenient to represent the stretched exponential function as a Prony series [53]-[54], i.e., as a discrete sum of simple exponential terms:

$$\exp(-x^\beta) \approx \sum_{i=1}^{N} w_i \exp(-K_i x), \tag{3}$$

where $x \equiv t/\tau$ and the weighting factors $w_i$ satisfy

$$\sum_{i=1}^{N} w_i = 1. \tag{4}$$

Use of a Prony series approximation to the stretched exponential function greatly increases the computational efficiency of the relaxation model, since the simple exponential terms in the Prony series allow for analytical integration of the relaxation equations. In this work, we perform a mathematical analysis of the Prony series representation of the stretched exponential function, including optimization of the set of positive $\{w_i, K_i\}$ coefficients as a function of $N$, the number of terms in the Prony series. While the optimization procedure is applicable to any value of the stretching exponent, $\beta$, we focus our attention on the specific cases of $\beta = \{3/7, 1/2, 3/5\}$, which are of particular interest in the context of the Phillips diffusion-trap model with $d^* = \{3/2, 2, 3\}$. We analyze the quality of the optimized Prony series fits in terms of both its representation of the stretched exponential decay function itself and also its first derivative. An inherent shortcoming of the Prony series representation is that it cannot reproduce the divergent slope of the stretched exponential function in the zero-time limit. Finally, we perform a frequency-domain analysis of the Prony series fits compared to the stretched exponential function and comment on some of the physical implications of using a Prony series approximation to stretched exponential decay.

## II. Modeling of Glass Relaxation

The nonequilibrium nature of the glassy state means that the usual set of thermodynamic state variables is insufficient for describing a glassy system, and additional order parameters are required to capture its full thermal history dependence [55]-[59]. As a first-order approximation, the nonequilibrium state of a glass can be described in terms of a fictive temperature, $T_f$, which represents the temperature at which the glass would be in equilibrium if the physical temperature of the system were instantaneously brought to $T = T_f$ [60]. This simplistic single-order parameter description of the nonequilibrium state of a glass was found to be insufficient for describing the detailed relaxation behavior of the glass, e.g., in the classic crossover experiments of Ritland [61], who showed that two glasses having the same composition and the same fictive temperature, but prepared via different thermal paths, can exhibit markedly different relaxation behaviors when held isothermally at $T = T_f$. Note that in the soft matter community this crossover effect is often attributed to Kovacs, who performed similar experiments on polymeric systems [62]-[63]. However, the work of Ritland [61] in the inorganic glass community preceded that of Kovacs [63] by seven years, so the crossover effect should be more properly attributed to Ritland.

To overcome this deficiency of a single fictive temperature description of the glassy state, Narayanaswamy [64] proposed describing the nonequilibrium state of glass in terms of multiple fictive temperatures, $\{T_{fi}\}$. For glass relaxation following stretched exponential decay, we may think of $\{T_{fi}\}$ as a set of fictive temperature components, where the average fictive temperature is

$$T_f = \sum_{i=1}^{N} w_i T_{fi}, \tag{5}$$

and the weighting coefficients, $w_i$, are defined by the Prony series approximation of the stretched exponential function, as in Eq. (3). In other words, there is one fictive temperature component, $T_{fi}$, associated with each individual term of the Prony series. The relaxation behavior of the glass can then be calculated in terms of relaxation of the individual fictive temperature components using a set of $N$ coupled first-order differential equations [53]:

$$\frac{dT_{fi}}{dt} = \frac{T(t) - T_{fi}(t)}{\tau_i[T(t), T_f(t)]}, \quad i = \{1, \ldots, N\}. \tag{6}$$

Here, $T(t)$ is the thermal history of the system, and the time constants $\{\tau_i\}$ are calculated from

$$\tau_i(T, T_f) = \tau(T, T_f) / K_i. \tag{7}$$

In other words, all $N$ terms share a common factor, $\tau$, which is a function of both the physical temperature, $T$, and the average fictive temperature, $T_f$, given by Eq. (5). The values of $\tau$ are scaled by the $K_i$ coefficients from the Prony series approximation in Eq. (3). Several models for $\tau(T,T_f)$ exist, including the classic models of Narayanaswamy [64] and Mazurin et al. [65] More recently, the Mauro-Allan-Potuzak (MAP) model for $\tau(T,T_f)$ has been shown to provide significant advantage in describing both the temperature and fictive temperature dependence of the relaxation time using a unified set of parameters [66]. The MAP model is also able to predict the composition dependence of $\tau(T,T_f)$ based on the underlying topology of the glass network in terms of temperature-dependent constraint theory [67]-[70]. The detailed form of the MAP model is provided in Ref. [70]. However, please note that the analysis of the Prony series presented in the current paper does not depend upon the particular choice of model for $\tau(T,T_f)$.

## III. Optimization of the Prony Series

Previously, Johnston has published a comprehensive analysis of stretched exponential relaxation expressed in terms of a continuous sum of simple exponential decays [71]. Unfortunately, no such thorough analysis exists for a discrete sum of simple exponentials, i.e., a Prony series, which is the relevant case for modeling of glass relaxation using a finite set of fictive temperature components, as in Eqs. (5)-(7). In this work, we find optimum values of the $\{w_i, K_i\}$ coefficients in Eq. (3) as a function of the number of terms in the Prony series, $N$, beginning with a single term representing simple exponential decay ($N = 1$) and continuing through $N = 12$. For a Prony series with $N$ terms, there are only $2N - 1$ free parameters, since the normalization condition of Eq. (4) must be satisfied.

Optimization in the case of $N = 1$ is straightforward, since there is only one free parameter, which can be determined by minimization of the root-mean-square error (RMSE) in the normalized time domain, $x$. However, for larger values of $N$ the optimization process is complicated by the presence of local minima in the RMSE phase space. To circumvent this problem, we adopt a combined stochastic/deterministic approach for optimization where a Monte Carlo approach is used to select an initial set of $\{w_i, K_i\}$ coefficient values, and then the RMSE is minimized deterministically following the Levenberg-Marquardt nonlinear least-squares method [72]. The number of random initial conditions sampled by the Monte Carlo routine increases geometrically with $N$ in order to account for the increased dimensionality of the phase space. In all cases, the RMSE is calculated over the domain of $x = [0, 40]$. This wide range of the normalized time, $x$, is chosen to capture the fat tail behavior of the stretched exponential relaxation function.

Optimized parameters for several particular scenarios of interest are provided in Tables I and II. In Table I, the optimized set of $\{w_i, K_i\}$ coefficients for $N = 8$ is given for three critical values of $\beta = \{3/5, 1/2, 3/7\}$ predicted from Phillips diffusion-trap theory. In Table II, the same values of the stretching exponent are considered, but increasing the number of terms in the Prony series to $N = 12$.

## IV. Time Domain Analysis

The optimized Prony series coefficients for $N = \{8, 12\}$ are plotted in Figure 1 for $\beta = \{3/5, 1/2, 3/7\}$. This figure shows that for all three values of $\beta$, the optimized terms of the Prony series are relatively evenly distributed over the $\log(K_i)$ space. The weighting factors, $w_i$, peak slightly below $K_i = 1$ and are asymmetrically distributed in $\log(K_i)$ space, with a greater weighting on the high $K_i$ side of the distribution to accurately capture the short-time scaling of the stretched exponential function, where the relaxation occurs most rapidly and shows the greatest discrepancy compared to simple exponential decay. As $\beta$ decreases, the stretched exponential decay function becomes increasing nonexponential, and hence there is a broader distribution of the $w_i$ coefficients in $\log(K_i)$ space.

The convergence and improved fitting quality of the Prony series with increasing $N$ is shown in Figure 2. The greatest difference between the stretched exponential function and the Prony series approximation occurs at short times, where the stretched exponential exhibits much faster decay than can be captured with a simple exponential function ($N = 1$). Simple exponential decay also fails to capture the fat tail of the stretched exponential at long times; however, this error asymptotically decreases to zero in the limit of infinite time. As the number of terms in the Prony series increases, a much closer approximation to the stretched exponential function can be obtained. As already mentioned, the terms in the Prony series are asymmetrically weighted toward higher values of $K_i$ to capture the short-time behavior and minimize the RMSE compared to the stretched exponential function. The terms in the Prony series having lower values of $K_i$ are necessary to capture the fat tail scaling of the stretched exponential decay.

The RMSE over the full range of normalized times, $x = [0, 40]$, is plotted in Figure 3(a) for $\beta = \{3/5, 1/2, 3/7\}$ and $N = [1, 12]$. Over this range of $\beta$ values, the RMSE increases with decreasing $\beta$ since the distribution is becoming increasingly nonexponential. In the context of the Phillips diffusion-trap model, these lower values of $\beta$ correspond to a lower effective dimensionality of the activated relaxation pathways. Figure 3(a) shows that the RMSE decreases exponentially with increasing $N$. The slope of $d\log(\text{RMSE})/dN$ is independent of the particular value of $\beta$ for this range of stretching exponents. This exponential scaling of the RMSE with $N$ can be used to calculate the minimum number of terms required to achieve a target level of accuracy with the Prony series fit. For example, if a ppm level of accuracy is desired, a minimum of 10 terms must be included for $\beta = 3/5$, whereas a minimum of 11 terms is required for the two lower values of $\beta = \{3/7, 1/2\}$.

In Figure 3(b) we again plot the RMSE calculated over $x = [0, 40]$, but now considering the full range of potential $\beta$ values from zero to one. The RMSE is plotted on a linear scale for both $N = 4$ (left axis) and $N = 8$ (right axis). The error vanishes in both the limits of $\beta \to 0$ and $\beta \to 1$, since in both of these limits the stretched exponential function reduces to a simple exponential. The maximum in RMSE, and hence the maximum in nonexponentiality, occurs around $\beta \approx 0.3$. Since all of the critical exponent values from the Phillips diffusion-trap model have $\beta > 0.3$, for any realistic glass relaxation process the level of nonexponentiality increases with decreasing $\beta$. In other words, for this realistic range of $\beta$ values, the quality of the Prony series fit becomes worse with decreasing $\beta$. It is worth noting that over our range of particular interest, $\beta = [3/7, 3/5]$, the scaling of the RMSE with $\beta$ is approximately linear.

Next, let us consider the slope of the stretched exponential decay in time. The first derivative of the stretched exponential function is

$$\frac{dg}{dt} = -\frac{\beta}{t}\left(\frac{t}{\tau}\right)^{\beta} \exp\left[-(t/\tau)^{\beta}\right], \tag{8}$$

which diverges in the limit of zero time for any $\beta < 1$:

$$\lim_{t \to 0} \frac{dg}{dt} = -\infty. \tag{9}$$

In contrast, the first derivative of the Prony series is

$$\frac{dg_{\text{Prony}}}{dt} = \frac{d}{dt}\sum_{i=1}^{N} w_i \exp(-K_i t/\tau) = -\frac{1}{\tau}\sum_{i=1}^{N} w_i K_i \exp(-K_i t/\tau). \tag{10}$$

At initial time, $t = 0$, the first derivative of the Prony series is finite, regardless of the number of terms in the series:

$$\lim_{t \to 0} \frac{dg_{\text{Prony}}}{dt} = -\frac{1}{\tau}\sum_{i=1}^{N} w_i K_i. \tag{11}$$

This difference represents a fundamental inconsistency between the stretched exponential function and the Prony series approximation, since it is impossible for a Prony series with any set of finite coefficients $\{w_i, K_i\}$ to reproduce the divergence in slope at initial time as exhibited by stretched exponential decay.

The slope of the relaxation function is plotted in Figure 4 for stretched exponential decay with $\beta = 3/7$ and Prony series fits with $N = \{1, 2, 4, 8\}$. The slope is convergent for all values of $N$ but becomes increasingly negative with a greater number of terms, approaching but never achieving the limit of the stretched exponential slope in Eq. (9). The error in the slope of the Prony series approximation is hence always divergent in the limit of $\{t, x\} \to 0$. The error evolves non-monotonically, exhibiting a local peak at $0 < x < 1$ before asymptotically approaching zero in the limit of $\{t, x\} \to \infty$.

## V. Frequency Domain Analysis

The first frequency domain analysis of relaxation using the stretched exponential function was conducted by Williams and Watts in their highly influential 1970 paper [73]. The Williams-Watts paper is largely responsible for popularizing the use of the stretched exponential function originally introduced by Kohlrausch more than a century previously. Their work was so influential that the stretched exponential function is also commonly known as the Kohlrausch-Williams-Watts (KWW) function [35]. Relaxation in the frequency domain has been studied extensively [74]-[78], although not within the context of a discrete Prony series.

Unfortunately, there is no exact analytical form for the Fourier transform of the stretched exponential decay function. In contrast, the Fourier transform of the Prony series is analytically soluble:

$$G_{\text{Prony}}(\omega) = \int_{-\infty}^{\infty} g_{\text{Prony}}(x) \exp(-j\omega x) dx . \qquad (12)$$

Since the relaxation function is defined only for positive time, we have

$$G_{\text{Prony}}(\omega) = \sum_{i=1}^{N} w_i \int_0^\infty \exp[-(K_i + j\omega)x] dx. \quad (13)$$

By substitution of variables, $u = -(K_i + j\omega)x$, we obtain

$$G_{\text{Prony}}(\omega) = -\sum_{i=1}^{N} \frac{w_i}{K_i + j\omega} \int_0^{-\infty} e^u du. \quad (14)$$

Solving the integral, we obtain the final expression for the Fourier transform of the Prony series:

$$G_{\text{Prony}}(\omega) = \sum_{i=1}^{N} \frac{w_i}{K_i + j\omega}. \quad (15)$$

Figure 5 plots the real and imaginary parts of Eq. (15) for optimized Prony series with $N = 1, 2, 4$, and 8, each fit to stretched exponential decay with $\beta = 3/7$. Both the (even) real and (odd) imaginary parts of the spectrum quickly converge with increasing $N$, providing a good approximation of the Fourier transform of the stretched exponential function, which is calculated numerically using a fast Fourier transform (FFT) algorithm [79].

The dependence of the stretched exponential spectrum on $\beta$ can be seen in Figure 6, where the real and imaginary parts of the spectra are plotted for $\beta = \{3/7, 1/2, 3/5\}$. The relaxation function has a greater spectral width for higher values of $\beta$, as can also be seen by the power spectral densities plotted in Figure 7(a). The spectral width decreases with lower values of $\beta$, since a narrower range of frequencies is required to produce the more abrupt initial decay at these lower values of the stretching exponent. Figure 7(b) shows the convergence of the power spectral density of the Prony series with increases $N$. Even with a small number of terms, the Prony series provides an excellent description of the power spectral density for the stretched exponential function. This indicates the analytical solution of the Fourier transform of the Prony series (Eq. (15)) may be used as a suitable substitute for the Fourier transform of the stretched exponential function, since this would otherwise need to be calculated numerically.

## VI. Discussion

The glassy state is often defined by what it is *not* as much as by what it is. This is encapsulated in several "non"s that describe what it means to be a glass [20]. First, glasses are *non-crystalline* materials lacking in the long-range periodic order characteristic of crystals. Second, glasses are thermodynamically *nonequilibrium* materials, continuously relaxing to the supercooled liquid state and having the ultimate fate of crystallizing [14]. Third, glasses are *nonergodic* since the relaxation time scales are long compared to the experimental observation time scale [1]-[2]. Finally, the shape of the relaxation function is *nonexponential*, described by the stretched exponential decay function [33].

For more than a century, the stretched exponential function was considered to be purely empirical, where the dimensionless stretching exponent $\beta$ is treated as a fitting parameter. However, the development of the Phillips diffusion-trap model has shown that stretched exponential relaxation is not only physically meaningful, but is also an inherent property of homogeneous glassy systems [33]. The derivation of the stretching exponent in terms of the topology of the relaxation pathways has revolutionized the understanding of this ubiquitous behavior, such that $\beta$ should no longer be treated as merely an empirical fitting parameter. This is important for both understanding the fundamental physics of glass relaxation behavior and for its practical impact in the high-tech glass industry [20].

Despite these advances in understanding the physics of stretched exponential relaxation, it is still computationally desirable to represent the stretched exponential function as a Prony series of weighted simple exponential decays operating over a range of time scales. This approach yields a description of glass relaxation in terms of a system of $N$ coupled differential equations, one for each fictive temperature component, $T_{fi}$, following Eq. (6). While a minimum of $N = 2$ is required to capture the qualitative features of the Ritland crossover effect, a greater number of terms is required to provide a quantitatively accurate description of the relaxation process. The number of terms that should be used depends on the desired accuracy of the model. The exponential decrease of RMSE with $N$, as depicted in Fig. 3(a), can be used to guide the choice of $N$ to achieve a target level of accuracy.

Despite the quantitative accuracy of the Prony series representation—with sufficiently large $N$—in both the time and frequency domains, it is not able to produce a divergent first-derivation of the relaxation function at initial time. This shortcoming, as well as the fact that the choice of $N$ is somewhat arbitrary, indicates that one should not view the individual terms of the Prony series as a realistic description of the actual relaxation modes in the glass. While it is tempting to ascribe physical or chemical meaning to these terms, we must keep in mind that the set of $\{w_i, K_i\}$ parameters are not independent, but rather defined by the single value of the stretching exponent, $\beta$. The physical meaning of the relaxation function derives from $\beta$, not from the set of $\{w_i, K_i\}$. Moreover, the description of a glass in terms of a plurality of fictive temperature components, $T_{fi}$, has been shown to be just a mathematical approximation and not capable of capturing the detailed nonequilibrium physics of the glassy state [58]-[59]. Previous work by Mauro et al. [58] has shown that even a continuum of fictive temperatures is incapable of capturing the true nonequilibrium state of a glass. This means that even a continuum of simple exponential decays, such as considered by Johnston [71], is insufficient for describing the true relaxation function of a glass. The main physical implication is that the glassy state, in general, cannot be fully described in terms of any combination of supercooled liquid states (i.e., even a continuum of such states). Mathematically, this means the locus of all supercooled liquid states do not form a complete basis set. The glassy state is unique from any supercooled liquid and can only be *approximated* with such a description. This is why a Prony series description of the true stretched exponential decay should be interpreted only as a convenient mathematical approximation, and not as the true underlying physics of the glassy state.

A remaining challenge is how best to capture the temperature dependence of the stretching exponent, $\beta(T)$, since the Phillips diffusion-trap model gives only the lower limit of $\beta$ at temperatures sufficiently below the glass transition. As the temperature increases, the value of $\beta$ increases monotonically until asymptotically approaching the upper limit of 1 in the ergodic liquid state [80]-[82]. However, currently there is no analytically derived model for the shape of $\beta(T)$. In the future, it would be highly desirable to incorporate temperature dependence in the Phillips diffusion-trap model to capture the full $\beta(T)$ curve, much as temperature-dependent constraint theory has generalized the Phillips topological constraint description of glass and liquid networks [83]-[89]. For now, it is worth considering the practical example of a liquid initially in equilibrium that is rate-cooled through the glass transformation regime. During this cooling process, the value of $\beta$ will decrease from unity to the lower limiting value predicted by the Phillips diffusion-trap model. Initially at high temperatures, the relaxation function can be described by a single exponential decay term. As the temperature is lowered, additional terms are required in the Prony series to provide an accurate description of the relaxation function. This can be seen in Fig. 3(b), where the RMSE of the Prony series fit is plotted as a function of $\beta$. From a practical modeling point of view, when selecting the number of terms used in the Prony series, one should consider the value of $\beta$ corresponding to the highest RMSE for the Prony series fit, viz., the lower limiting value from the Phillips diffusion-trap model. This value should therefore govern the choice of $N$ for the entire simulation.

**VII. Conclusions**

Stretched exponential relaxation is a universal feature of homogeneous glassy systems. Practical implementations of glass relaxation models typically rely on approximating the stretched exponential function as a Prony series with a finite number of terms, where each term in the Prony series has an associated fictive temperature component. The fictive temperature components each relax toward the instantaneous value of the physical temperature following a set of coupled first-order differential equations. Accurate modeling of the glass relaxation process therefore relies upon obtaining optimized coefficients for the Prony series.

Following a combined Monte Carlo/Levenberg-Marquardt optimization routine, we have determined the optimized coefficients of the Prony series as functions of both the stretching exponent, $\beta$, and the number of terms in the Prony series, $N$. Our results show that the RMSE of the optimized Prony series fit decreases exponentially with increasing $N$. This scaling behavior can be used to determine the number of terms required to achieve a particular level of accuracy when modeling the relaxation behavior of glass. The RMSE vanishes in the limits of both $\beta \to 0$ and $\beta \to 1$, where the stretched exponential function reduces to a simple exponential.

While a Prony series fit can give a highly accurate representation of stretched exponential decay in both the time and frequency domains, it is not able to reproduce the divergence of the stretched exponential slope in the limit of zero time. Moreover, caution should be exercised when ascribing physical meaning to the individual terms of the Prony series, e.g., by viewing each fictive temperature component as a physically meaningful relaxation mode in the glass, individually following a simple exponential decay. Rather, the Prony series should be interpreted as a mathematically convenient approximation to the physically meaningful stretched exponential function, which arises from the statistics of all relaxation pathways in a homogeneous glass.

**Acknowledgement:** We are grateful for many enlightening discussions with Douglas C. Allan.

**TABLES**

**Table I.** Parameters of the optimized Prony series fits with $N = 8$ for stretched exponential relaxation with $\beta = 3/7$, $1/2$, and $3/5$.

|   | $\beta = 3/7$ | | $\beta = 1/2$ | | $\beta = 3/5$ | |
|---|---|---|---|---|---|---|
| $i$ | $w_i$ | $K_i$ | $w_i$ | $K_i$ | $w_i$ | $K_i$ |
| 1 | 0.05014 | 0.04812 | 0.03747 | 0.07735 | 0.03417 | 0.14633 |
| 2 | 0.14755 | 0.14694 | 0.15374 | 0.19035 | 0.18347 | 0.28920 |
| 3 | 0.18607 | 0.39937 | 0.21791 | 0.46310 | 0.26687 | 0.61223 |
| 4 | 0.17905 | 1.07652 | 0.20267 | 1.17279 | 0.21782 | 1.42012 |
| 5 | 0.14976 | 3.08783 | 0.15447 | 3.21677 | 0.13987 | 3.65392 |
| 6 | 0.11362 | 9.91235 | 0.10528 | 9.95212 | 0.08121 | 10.69645 |
| 7 | 0.08050 | 38.09496 | 0.06676 | 37.10866 | 0.04448 | 37.81238 |
| 8 | 0.09331 | 254.89688 | 0.06168 | 236.28861 | 0.03209 | 221.25717 |

**Table II.** Parameters of the optimized Prony series fits with $N = 12$ for stretched exponential relaxation with $\beta = 3/7$, $1/2$, and $3/5$.

|   | $\beta = 3/7$ | | $\beta = 1/2$ | | $\beta = 3/5$ | |
|---|---|---|---|---|---|---|
| $i$ | $w_i$ | $K_i$ | $w_i$ | $K_i$ | $w_i$ | $K_i$ |
| 1 | 0.02792 | 0.03816 | 0.01694 | 0.06265 | 0.01043 | 0.12022 |
| 2 | 0.09567 | 0.10117 | 0.08574 | 0.13381 | 0.08117 | 0.20610 |
| 3 | 0.13049 | 0.22822 | 0.14468 | 0.26816 | 0.17168 | 0.35680 |
| 4 | 0.13388 | 0.47142 | 0.15870 | 0.52050 | 0.19624 | 0.63293 |
| 5 | 0.12456 | 0.94243 | 0.14514 | 1.00410 | 0.16742 | 1.15481 |
| 6 | 0.10976 | 1.88828 | 0.12095 | 1.96395 | 0.12467 | 2.17404 |
| 7 | 0.09256 | 3.86312 | 0.09512 | 3.94401 | 0.08711 | 4.23811 |
| 8 | 0.07525 | 8.15604 | 0.07188 | 8.20241 | 0.05896 | 8.59467 |
| 9 | 0.05938 | 17.92388 | 0.05275 | 17.81155 | 0.03913 | 18.25401 |
| 10 | 0.04587 | 41.47225 | 0.03791 | 40.85894 | 0.02559 | 41.07522 |
| 11 | 0.03588 | 104.13591 | 0.02749 | 102.07104 | 0.01688 | 100.99297 |
| 12 | 0.06877 | 402.71691 | 0.04270 | 383.52267 | 0.02071 | 363.84147 |

**FIGURES**

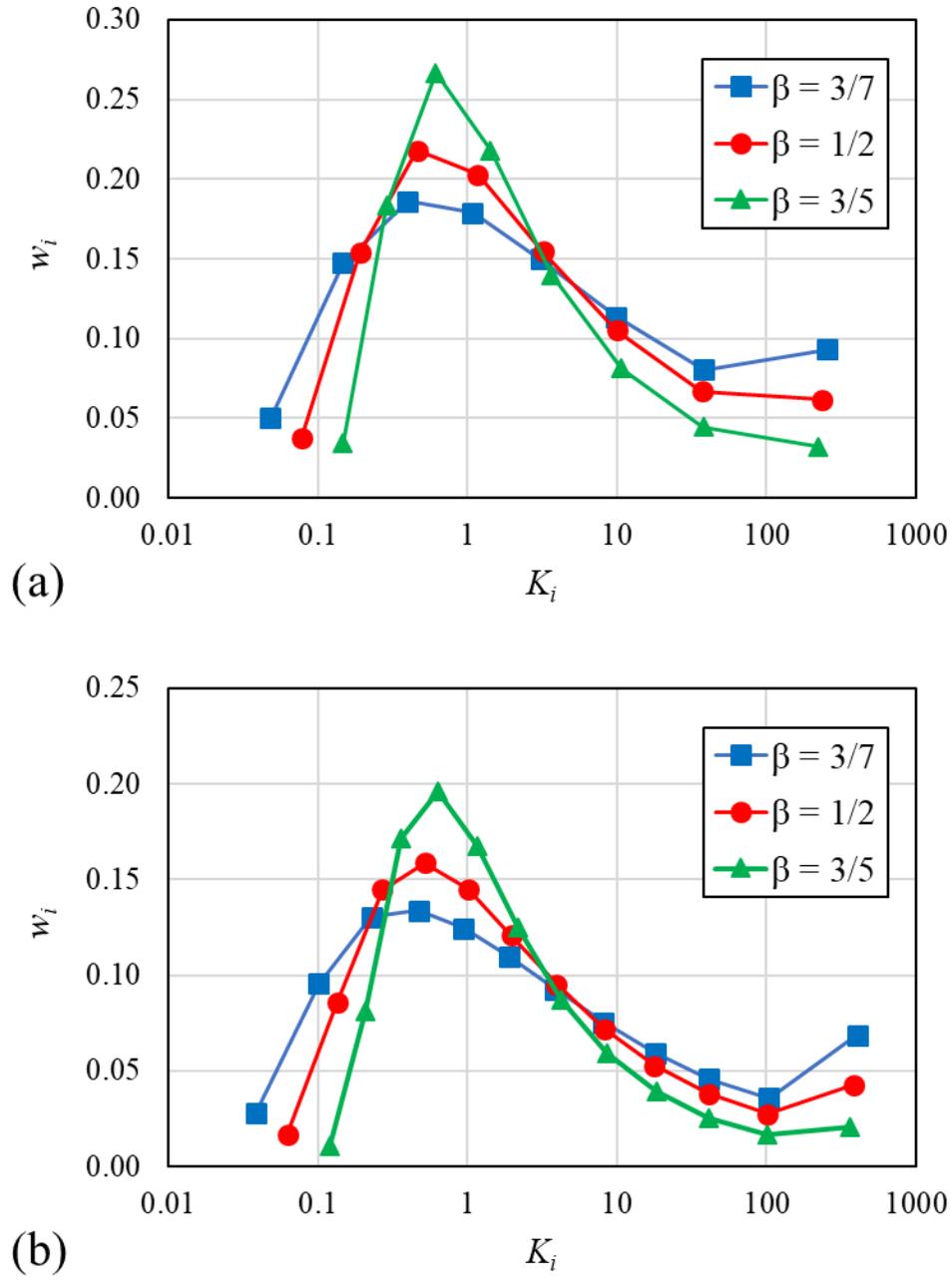

**Figure 1.** Optimized parameters of the Prony series fits for stretched exponential relaxation with $\beta = 3/7$, $1/2$, and $3/5$ with (a) $N = 8$ and (b) $N = 12$.

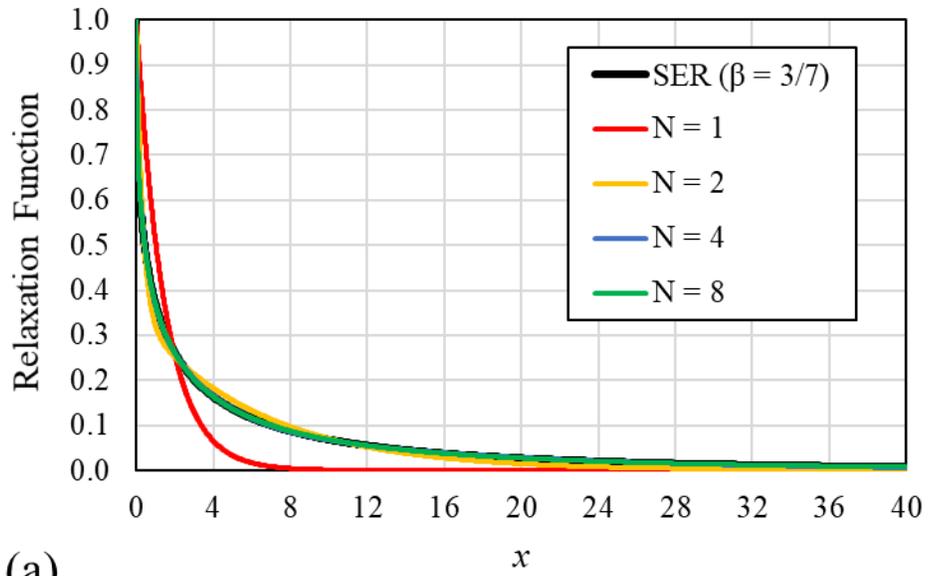

(a)

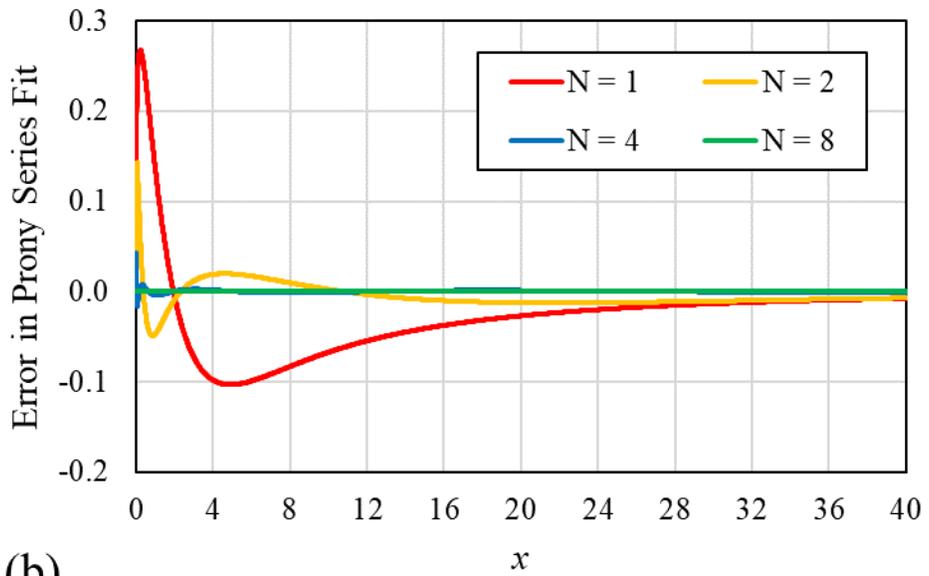

(b)

**Figure 2.** (a) Stretched exponential relaxation with $\beta = 3/7$ versus optimized Prony series fits with $N = 1, 2, 4$, and 8. (b) Error in the Prony series fits as a function of normalized time, $x$.

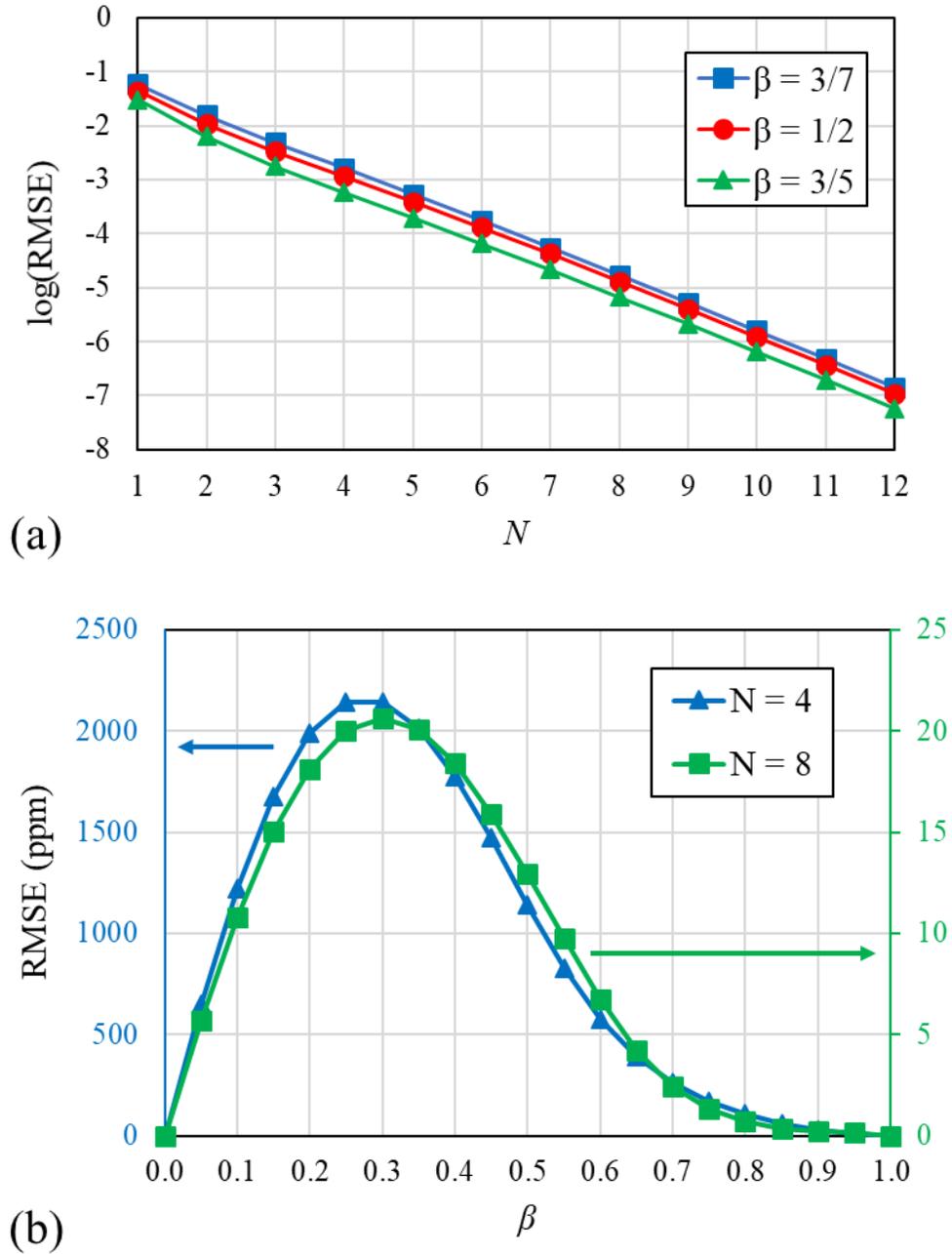

**Figure 3.** (a) Root-mean-square error (RMSE) of optimized Prony series fits to stretched exponential relaxation (SER) with $\beta$ = 3/5, 1/2, and 3/7 as a function of number of terms in the Prony series, $N$ = [1, 12]. (b) RMSE of optimized Prony series fits to SER with $N$ = {4, 8} and $\beta$ = [0, 1].

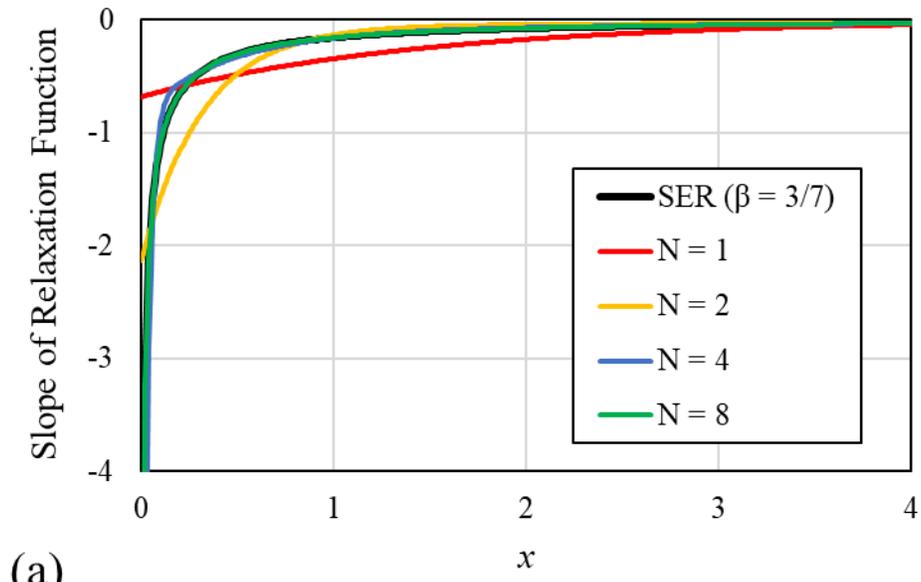

(a)

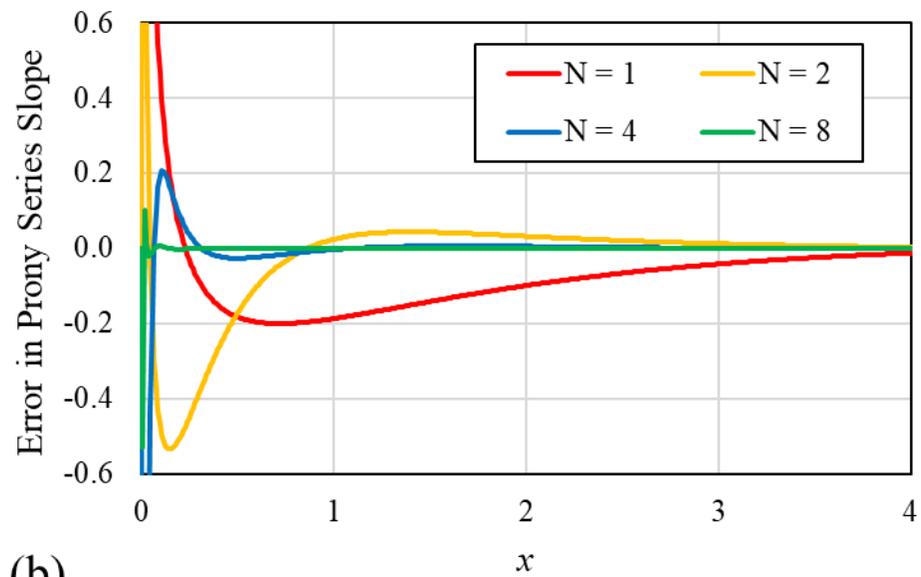

(b)

**Figure 4.** (a) First derivative of stretched exponential relaxation (SER) with $\beta = 3/7$ versus optimized Prony series representations with $N = 1, 2, 4,$ and $8$. (b) Error in the first derivatives as a function of normalized time.

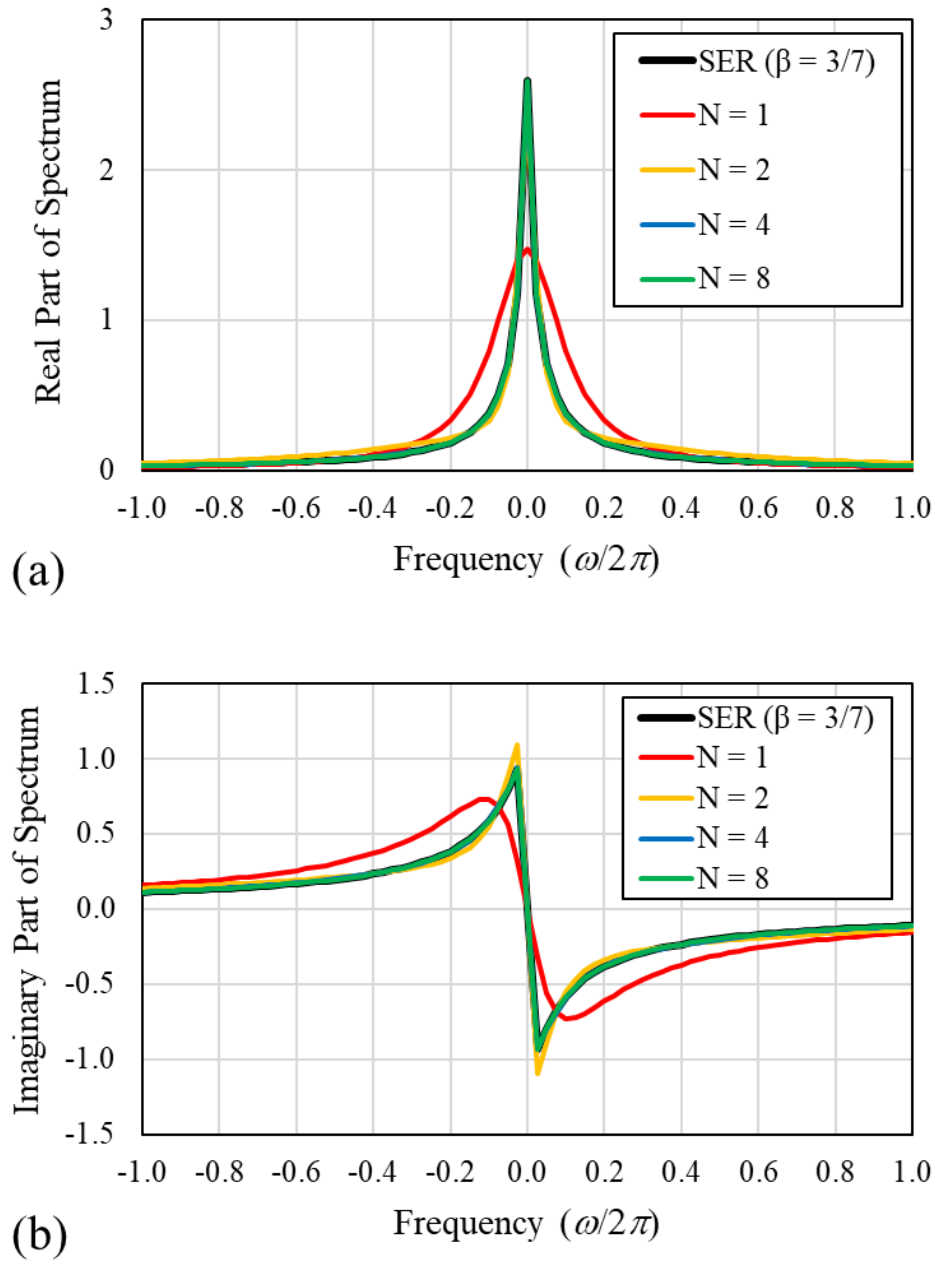

**Figure 5.** (a) Real and (b) imaginary part of the Fourier transform spectra of the stretched exponential function with $\beta = 3/7$ versus that of the optimized Prony series fits with $N = 1, 2, 4,$ and 8.

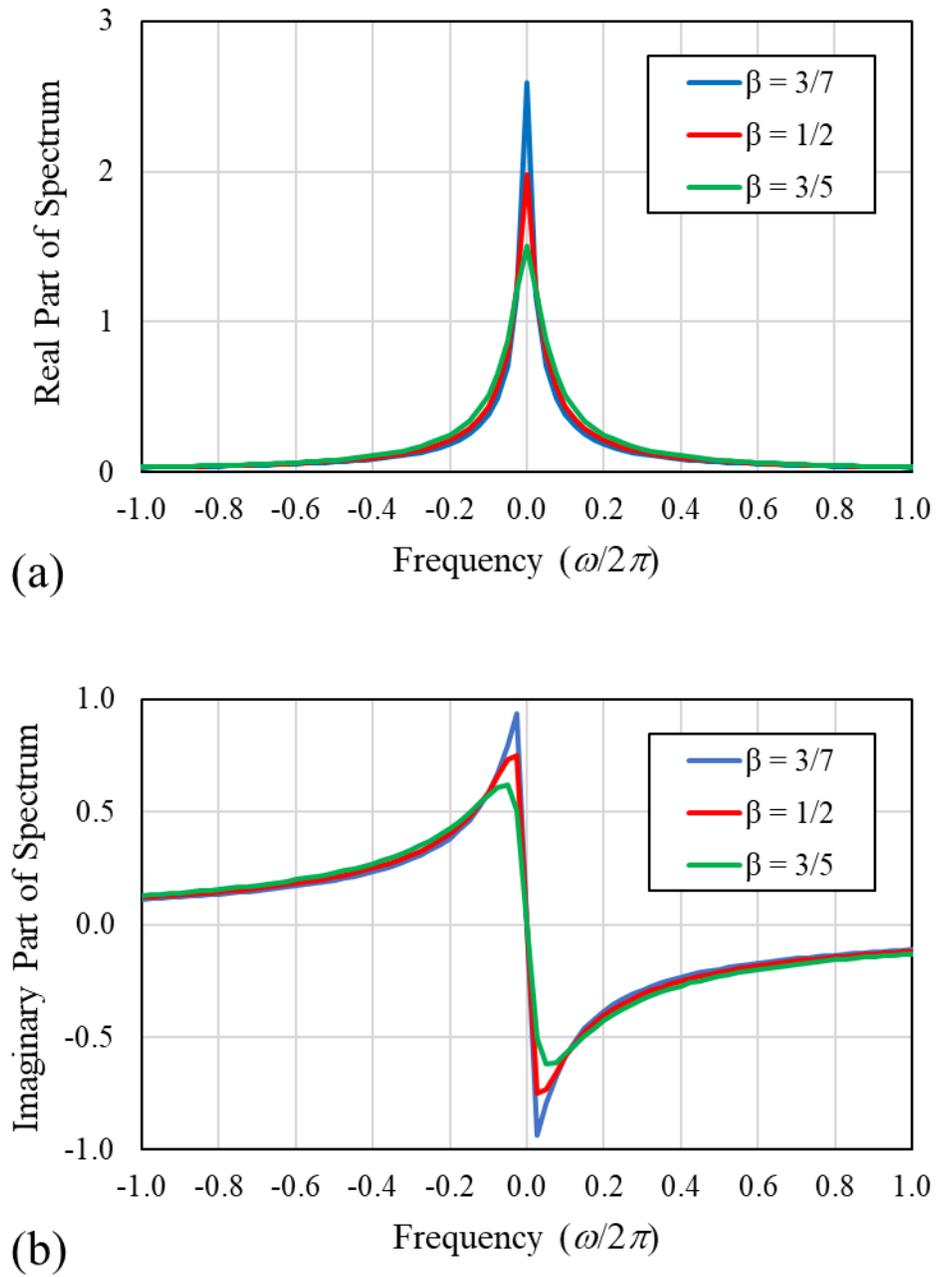

**Figure 6.** (a) Real and (b) imaginary part of the Fourier transform spectra of the stretched exponential function with $\beta$ = 3/7, 1/2, and 3/5.

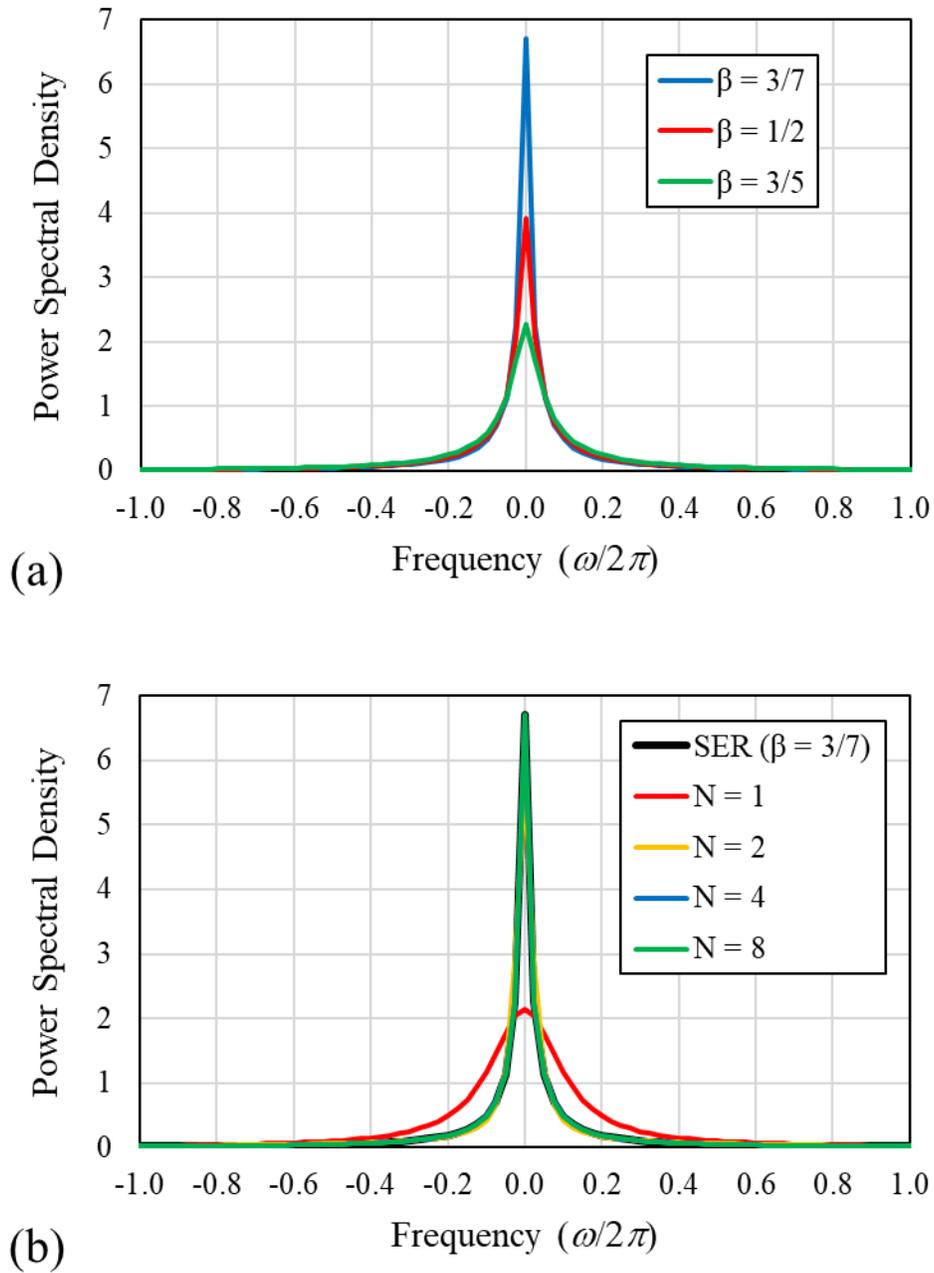

**Figure 7.** (a) Power spectral density of stretched exponential relaxation (SER) functions with $\beta$ = 3/7, 1/2, and 3/5. (b) Convergence of the power spectral density for Prony series fits of the SER function with $\beta$ = 3/7.